# Coupled Spin-Orbital Texture in a Prototypical Topological Insulator

Y. Cao[1], J.A. Waugh[1], N.C. Plumb[2], T.J. Reber[1], S. Parham[1], G. Landolt[2,3], Z. Xu[4], A. Yang[4], J. Schneeloch[4], G. Gu[4], J.H. Dil[2,3], D.S. Dessau[1,&]

1. Department of Physics, University of Colorado, Boulder, CO 80309, USA
2. Swiss Light Source, Paul Scherrer Institut, CH-5232 Villigen PSI, Switzerland
3. Physik-Institut, Universität Zürich, Winterthurerstrasse 190, CH-8057 Zürich, Switzerland
4. Condensed Matter Physics and Materials Science Department, Brookhaven National Labs, Upton, NY 11973, USA

**Abstract:** One of the most important properties of topological insulators (TIs) is the helical spin texture of the Dirac surface states, which has been theoretically and experimentally argued to be left-handed helical above the Dirac point and right handed helical below. However, since the spin is not a good quantum number in these strongly spin-orbit coupled systems, this can not be a complete statement, and we must consider the total angular momentum $\vec{J} = \vec{L} + \vec{S}$ that is a contribution of the spin and orbital terms. Using a combination of orbital and spin-resolved angle-resolved photoemission spectroscopy (ARPES), we show a direct link between the different orbital and spin components, with a "backwards" spin texture directly observed for the in-plane orbital states of $Bi_2Se_3$.

**Topological insulators** (TI's) have captivated the interest of the scientific community because of their novel properties and their potential for future technologies such as spintronics and quantum computation[1,2]. Their most famous property is a metallic "topologically protected" surface state enclosing an insulating bulk – the chocolate coating enclosing the ice-cream innards of a Klondike-brand bar. Similarly to graphene (another two dimensional metal) these surface states have a "Dirac-cone" band structure; the electronic energy-momentum dispersion relation is linear (massless) and meets at a single "Dirac point" (DP – see Figure 1). However, while the graphene Dirac cone has the usual pair of spin-degenerate states that any band respecting the Pauli principle would be expected to have, the novel state in the TI's contains only half as many electrons. These states are almost universally described as helically spin-polarized: left-handed above the so-called Dirac point (DP) where the energy-momentum dispersion relation touches at a point, switching to right-handed below the DP [1,2,3,4,5,6,7,8,9,10]. However, we note that the same spin-orbit coupling that is necessary for the creation of the topological state means that the spin is no longer a good quantum number; instead we must consider the total angular momentum $\vec{J} = \vec{L} + \vec{S}$, which can be thought of as a pseudospin. Here we show that the impact of this simple statement on the actual spin and orbital degrees of freedom is surprisingly profound.

Rather than the spin states $\vec{S}$, it is the pseudospin states $\vec{J}$ that should have the pure left or right-handed helicity as illustrated in figure 1a. On the other hand, $\vec{J}$ is not a measurable quantity and we can only measure or take advantage of the spin or orbital degrees of freedom. Discerning how these spin and orbital degrees of freedom interact with each other and together make up the conserved pseudospin quantity is clearly of fundamental importance for our understanding of the physics of topological insulators, and also may be relevant for applications which may hope to take advantage of their novel spin properties.

The first step towards deconvolving the spin and orbital degrees of freedom came with our "orbital-selective" ARPES (angle-resolved photoemission spectroscopy) measurements of the prototypical TI $Bi_2Se_3$ [11]. Linearly-polarized incident photons with either s or p polarization allowed us to separate out the relatively weak in-plane orbital states from the out-of-plane $p_z$ orbitals (schematically illustrated in Figure 1b) that had dominated previous ARPES measurements. A symmetry analysis of the data as well as our first-principles theoretical effort showed that the in-plane orbital states displayed an exotic texture, switching from tangential to the equal energy surface well above the DP to radial below it (Figure 1c) [11]. For states away from the DP, while other orbital terms also exist, these tangential orbitals are the dominant set of orbital states that make up the band structure of $Bi_2Se_3$. The question at hand is how these newly deconvolved orbital states should couple to the spin degree of freedom.

Stanford theorist Shoucheng Zhang and his group seized upon this result, doing two things [12]: first, they confirmed the details of the orbital texture switch with his own calculations and second, they predicted that the orbital texture should imply a novel spin texture component that is backwards from what had previously been observed, if only the spin texture of the in-plane orbitals could be separated out from the spin texture of the out-of-plane orbitals. More specifically they predicted the spin texture shown in Figures 1b and 1c: for the out-of-plane $p_z$ orbitals shown in figure 1b the spin texture is the standard texture that has been known for a number of years for the TIs [1,2,3,4,5,6,7,8,9,10**Error! Bookmark not defined.**]. On the other hand, for the in-plane orbitals they predicted a very unusual and unexpected spin texture, as shown in Figure 1c; it is right handed helical both above and below the DP, i.e. the spin texture associated is "backwards" from the normally observed left-handed spin texture above the DP. Our new experiments described here confirm this prediction.

Our new experiments combine spin-polarized electron detection with the orbital selective p and s polarization of the incident photons. The experiments were performed on single crystals of $Bi_2Se_3$ cleaved and measured in-vacuo at T=21K with a photon energy of 47 eV at the COPHEE endstation at the Swiss Light Source [13]. This facility utilizes a helical undulator, allowing us to change the incident photon polarization from linear s to linear p without altering any other parameters of the experiment, and the multi-axis spin detector at this facility allows for a deconvolution of the various spin degrees of freedom, as described elsewhere [14].

Panels a and b of Figure 2 show spin-integrated measurements taken with p and s polarization respectively, along the Γ-K cut through the Brillouin zone. Aside from the reduction in experimental statistics for the s-polarized data, these panels show an identical E vs. k dispersion for both polarizations. This is expected because both polarizations are probing the same surface Dirac band, but only projecting out the different orbital contributions that make up this band. On the other hand, the spin measurements (Figures 2c and 2d), taken at a fixed energy of -50 meV in the upper cone (along the black dashed lines of Figures 2a and 2b) are opposite for the two panels – they show a left-handed spin helicity for the p polarization data and a right-handed spin helicity for the s polarization data – a result fully consistent with Zhang's predictions [12] and the illustrations of Figure 1. No previous experiment has indicated the possibility that a right-handed spin texture may exist above the DP, so without Refs 11 and 12 this is a completely unexpected result. The magnitude of the spin polarization detected in the present experiment is between 50% and 100% for both polarizations, with the uncertainty a result of the background counts and the effective momentum resolution of the experiment, neither of which are known with complete precision. We note that our results leave open the possibility of a small in-plane radial or out-of-plane $k_z$ component of spin for the in-plane orbital states.

While the coupled spin-orbital texture observed here provides a novel insight into the nature of the topological surface state, it also must be relevant for any future applications which may wish to take advantage of these helically polarized surface states – which are only fully spin polarized if the different orbital contributions can be selectively accessed.

**Acknowledgements:** We thank Mike Hermele and Shoucheng Zhang for very helpful discussions. This work was supported by the US National Science Foundation under grant DMR1007014, by the US Department of Energy under Contract No. DE-AC03-76SF00098, and by the Swiss National Science Foundation.

[&] To whom correspondence should be addressed: Dessau@Colorado.edu

**Figure 1.** Schematic of the surface energy / momentum Dirac cone of $Bi_2Se_3$. (a) The left and right handed pseudospin vector J, which is presumed to be purely left handed in the upper cone and purely right-handed in the lower cone. The spin vectors S shown in the right panels are not pure but can be separated by a decomposition into the orbital wavefunctions. The two dominant orbital contributions are shown in panels (b) which feature out-of-plane $p_z$ orbitals observed with out-of-plane p-polarization and (c) in-plane orbitals observed with s-polarization. As originally shown in [11], the in-plane orbitals are predominantly tangential above the DP and radial below the DP. The measured spin polarization is illustrated by the red arrows in panels (b) and (c).

**Figure 2.** ARPES and spin-ARPES data of cleaved single crystals of $Bi_2Se_3$ with different photon polarization. Panels (a) and (b) show spin-integrated data along the Γ-K direction using the two photon polarizations, selecting out (a) the $p_z$ orbital and (b) the in-plane tangential and radial orbital contribution to the Dirac surface states. (c) and (d) show the measured spin asymmetry data taken at a constant energy cut of -50 meV (black line) in the upper cone. The measured spin polarization is of opposite helicity for the two experimental configurations.

Figure 1

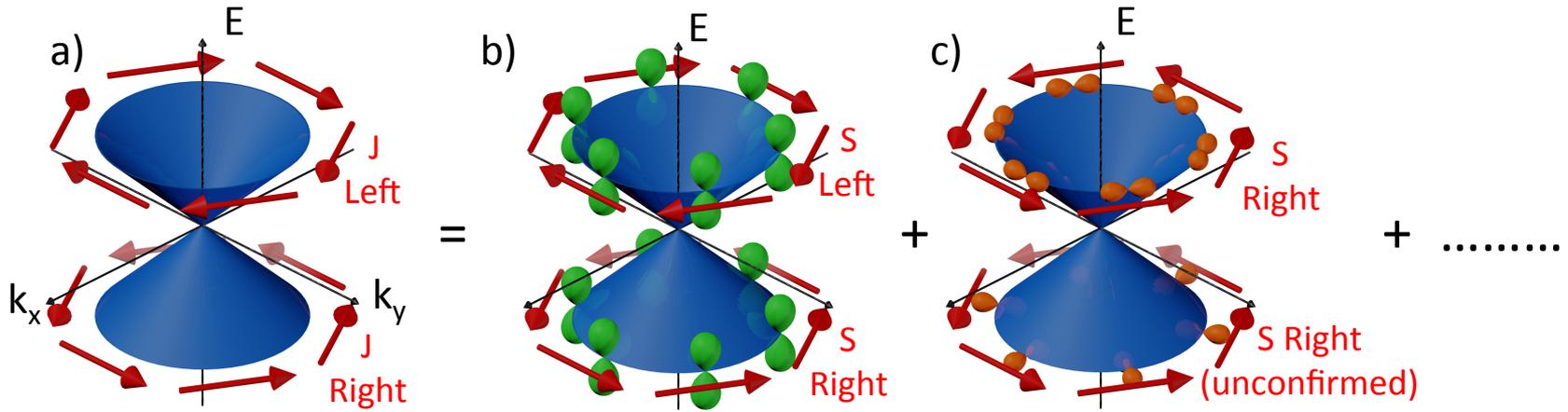

Figure 2

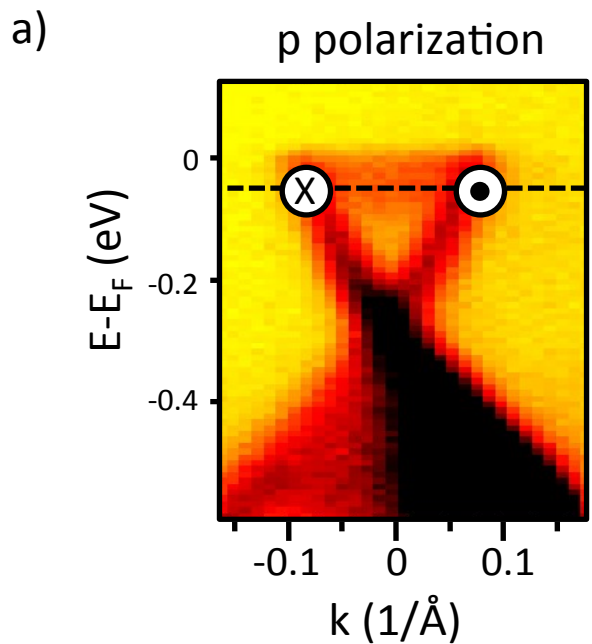 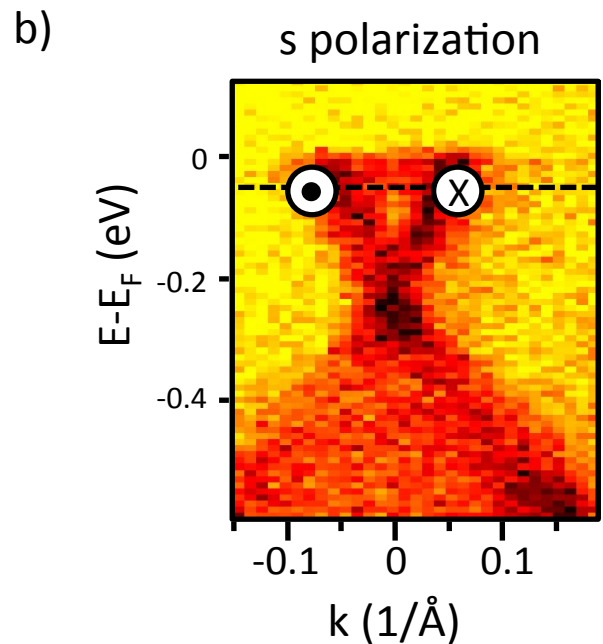
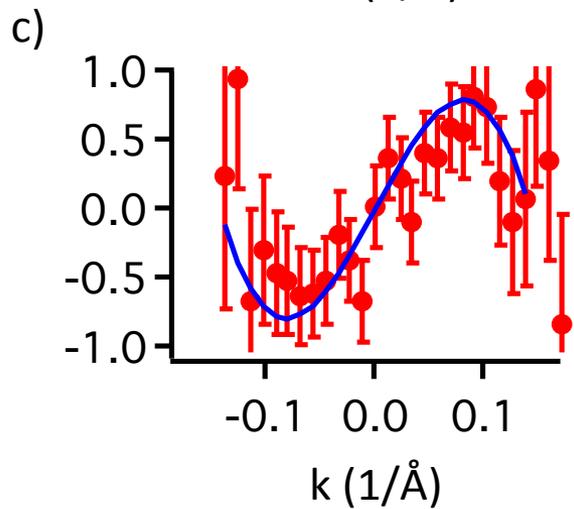 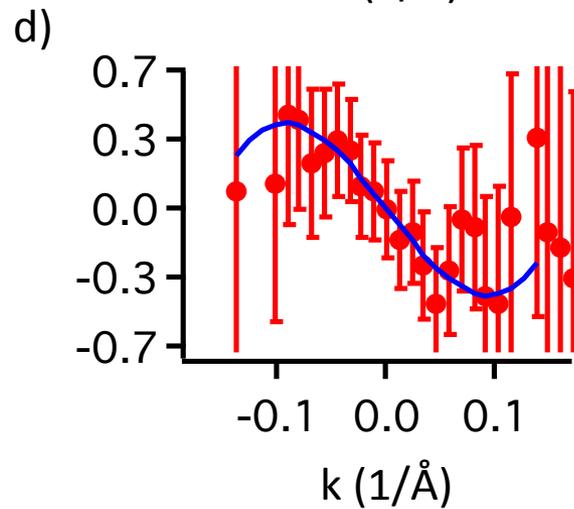